\pdfoutput=1
\documentclass[12pt,a4paper,titlepage]{article}
\usepackage[utf8]{inputenc}
\usepackage[top=50pt,bottom=50pt,left=68pt,right=66pt]{geometry}
\usepackage{amsmath}
\usepackage{booktabs}
\usepackage{amssymb}
\usepackage[title]{appendix}
\usepackage{mathrsfs}
\usepackage{float}
\usepackage{multirow}
\usepackage{graphicx,caption,subcaption}
\usepackage[space]{grffile}

\begin{document}
\renewcommand{\arraystretch}{1.3}

\makeatletter
\def\@hangfrom#1{\setbox\@tempboxa\hbox{{#1}}%
      \hangindent 0pt%\wd\@tempboxa
      \noindent\box\@tempboxa}
\makeatother

% Underline for text or math

\def\un#1{\relax\ifmmode\@@underline#1\else
        $\@@underline{\hbox{#1}}$\relax\fi}

% Accents and foreign (in text):

\let\under=\unt                 % bar-under (but see \un above)
\let\ced=\ce                    % cedilla
\let\du=\du                     % dot-under
\let\um=\Hu                     % Hungarian umlaut
\let\sll=\lp                    % slashed (suppressed) l (Polish)
\let\Sll=\Lp                    % " L
\let\slo=\os                    % slashed o (Scandinavian)
\let\Slo=\Os                    % " O
\let\tie=\ta                    % tie-after (semicircle connecting two letters)
\let\br=\ub                     % breve
                % Also: \`        grave
                %       \'        acute
                %       \v        hacek (check)
                %       \^        circumflex (hat)
                %       \~        tilde (squiggle)
                %       \=        macron (bar-over)
                %       \.        dot (over)
                %       \"        umlaut (dieresis)
                %       \aa \AA   A-with-circle (Scandinavian)
                %       \ae \AE   ligature (Latin & Scandinavian)
                %       \oe \OE   " (French)
                %       \ss       es-zet (German sharp s)
                %       \$  \#  \&  \%  \pounds  {\it\&}  \dots

% Abbreviations for Greek letters

\def\a{\alpha}
\def\b{\beta}
\def\c{\chi}
\def\d{\delta}
\def\e{\epsilon}
\def\f{\phi}
\def\g{\gamma}
\def\h{\eta}
\def\i{\iota}
\def\j{\psi}
\def\k{\kappa}
\def\l{\lambda}
\def\m{\mu}
\def\n{\nu}
\def\o{\omega}
\def\p{\pi}
\def\q{\theta}
\def\r{\rho}
\def\s{\sigma}
\def\t{\tau}
\def\u{\upsilon}
\def\x{\xi}
\def\z{\zeta}
\def\D{\Delta}
\def\F{\Phi}
\def\G{\Gamma}
\def\J{\Psi}
\def\L{\Lambda}
\def\O{\Omega}
\def\P{\Pi}
\def\Q{\Theta}
\def\S{\Sigma}
\def\U{\Upsilon}
\def\X{\Xi}

% Varletters

\def\ve{\varepsilon}
\def\vf{\varphi}
\def\vr{\varrho}
\def\vs{\varsigma}
\def\vq{\vartheta}

% Calligraphic letters

\def\ca{{\cal A}}
\def\cb{{\cal B}}
\def\cc{{\cal C}}
\def\cd{{\cal D}}
\def\ce{{\cal E}}
\def\cf{{\cal F}}
\def\cg{{\cal G}}
\def\ch{{\cal H}}
\def\ci{{\cal I}}
\def\cj{{\cal J}}
\def\ck{{\cal K}}
\def\cl{{\cal L}}
\def\cm{{\cal M}}
\def\cn{{\cal N}}
\def\co{{\cal O}}
\def\cp{{\cal P}}
\def\cq{{\cal Q}}
\def\car{{\cal R}}
\def\cs{{\cal S}}
\def\ct{{\cal T}}
\def\cu{{\cal U}}
\def\cv{{\cal V}}
\def\cw{{\cal W}}
\def\cx{{\cal X}}
\def\cy{{\cal Y}}
\def\cz{{\cal Z}}

% Fonts

\def\Sc#1{{\hbox{\sc #1}}}      % script for single characters in equations
\def\Sf#1{{\hbox{\sf #1}}}      % sans serif for single characters in equations

                        % Also:  \rm      Roman (default for text)
                        %        \bf      boldface
                        %        \it      italic
                        %        \mit     math italic (default for equations)
                        %        \sl      slanted
                        %        \em      emphatic
                        %        \tt      typewriter
                        % and sizes:    \tiny
                        %               \scriptsize
                        %               \footnotesize
                        %               \small
                        %               \normalsize
                        %               \large
                        %               \Large
                        %               \LARGE
                        %               \huge
                        %               \Huge

% Math symbols

\def\slpa{\slash{\pa}}                            % slashed partial derivative
\def\slin{\SLLash{\in}}                                   % slashed in-sign
\def\bo{{\raise-.3ex\hbox{\large$\Box$}}}               % D'Alembertian
\def\cbo{\Sc [}                                         % curly "
\def\pa{\partial}                                       % curly d
\def\de{\nabla}                                         % del
\def\dell{\bigtriangledown}                             % hi ho the dairy-o
\def\su{\sum}                                           % summation
\def\pr{\prod}                                          % product
\def\iff{\leftrightarrow}                               % <-->
\def\conj{{\hbox{\large *}}}                            % complex conjugate
\def\ltap{\raisebox{-.4ex}{\rlap{$\sim$}} \raisebox{.4ex}{$<$}}   % < or ~
\def\gtap{\raisebox{-.4ex}{\rlap{$\sim$}} \raisebox{.4ex}{$>$}}   % > or ~
\def\TH{{\raise.2ex\hbox{$\displaystyle \bigodot$}\mskip-4.7mu \llap H \;}}
\def\face{{\raise.2ex\hbox{$\displaystyle \bigodot$}\mskip-2.2mu \llap {$\ddot
        \smile$}}}                                      % happy face
\def\dg{\sp\dagger}                                     % hermitian conjugate
\def\ddg{\sp\ddagger}                                   % double dagger
                        % Also:  \int  \oint              integral, contour
                        %        \hbar                    h bar
                        %        \infty                   infinity
                        %        \sqrt                    square root
                        %        \pm  \mp                 plus or minus
                        %        \cdot  \cdots            centered dot(s)
                        %        \oplus  \otimes          group theory
                        %        \equiv                   equivalence
                        %        \sim                     ~
                        %        \approx                  approximately =
                        %        \propto                  funny alpha
                        %        \ne                      not =
                        %        \le \ge                  < or = , > or =
                        %        \{  \}                   braces
                        %        \to  \gets               -> , <-
                        % and spaces:  \,  \:  \;  \quad  \qquad
                        %              \!                 (negative)

\font\tenex=cmex10 scaled 1200

% Math stuff with one argument

\def\sp#1{{}^{#1}}                              % superscript (unaligned)
\def\sb#1{{}_{#1}}                              % sub"
\def\oldsl#1{\rlap/#1}                          % poor slash
\def\slash#1{\rlap{\hbox{$\mskip 1 mu /$}}#1}      % good slash for lower case
\def\Slash#1{\rlap{\hbox{$\mskip 3 mu /$}}#1}      % " upper
\def\SLash#1{\rlap{\hbox{$\mskip 4.5 mu /$}}#1}    % " fat stuff (e.g., M)
\def\SLLash#1{\rlap{\hbox{$\mskip 6 mu /$}}#1}      % slash for no-in sign
\def\PMMM#1{\rlap{\hbox{$\mskip 2 mu | $}}#1}   %
\def\PMM#1{\rlap{\hbox{$\mskip 4 mu ~ \mid $}}#1}       %
\def\Tilde#1{\widetilde{#1}}                    % big tilde
\def\Hat#1{\widehat{#1}}                        % big hat
\def\Bar#1{\overline{#1}}                       % big bar
\def\sbar#1{\stackrel{*}{\Bar{#1}}}             % big bar with star
\def\bra#1{\left\langle #1\right|}              % < |
\def\ket#1{\left| #1\right\rangle}              % | >
\def\VEV#1{\left\langle #1\right\rangle}        % < >
\def\abs#1{\left| #1\right|}                    % | |
\def\leftrightarrowfill{$\mathsurround=0pt \mathord\leftarrow \mkern-6mu
        \cleaders\hbox{$\mkern-2mu \mathord- \mkern-2mu$}\hfill
        \mkern-6mu \mathord\rightarrow$}
\def\dvec#1{\vbox{\ialign{##\crcr
        \leftrightarrowfill\crcr\noalign{\kern-1pt\nointerlineskip}
        $\hfil\displaystyle{#1}\hfil$\crcr}}}           % <--> accent
\def\dt#1{{\buildrel {\hbox{\LARGE .}} \over {#1}}}     % dot-over for sp/sb
\def\dtt#1{{\buildrel \bullet \over {#1}}}              % alternate "
\def\der#1{{\pa \over \pa {#1}}}                % partial derivative
\def\fder#1{{\d \over \d {#1}}}                 % functional derivative
                % Also math accents:    \bar
                %                       \check
                %                       \hat
                %                       \tilde
                %                       \acute
                %                       \grave
                %                       \breve
                %                       \dot    (over)
                %                       \ddot   (umlaut)
                %                       \vec    (vector)

% Math stuff with more than one argument

\def\frac#1#2{{\textstyle{#1\over\vphantom2\smash{\raise.20ex
        \hbox{$\scriptstyle{#2}$}}}}}                   % fraction
\def\half{\frac12}                                        % 1/2
\def\sfrac#1#2{{\vphantom1\smash{\lower.5ex\hbox{\small$#1$}}\over
        \vphantom1\smash{\raise.4ex\hbox{\small$#2$}}}} % alternate fraction
\def\bfrac#1#2{{\vphantom1\smash{\lower.5ex\hbox{$#1$}}\over
        \vphantom1\smash{\raise.3ex\hbox{$#2$}}}}       % "
\def\afrac#1#2{{\vphantom1\smash{\lower.5ex\hbox{$#1$}}\over#2}}    % "
\def\partder#1#2{{\partial #1\over\partial #2}}   % partial derivative of
\def\parvar#1#2{{\d #1\over \d #2}}               % variation of
\def\secder#1#2#3{{\partial^2 #1\over\partial #2 \partial #3}}  % second "
\def\on#1#2{\mathop{\null#2}\limits^{#1}}               % arbitrary accent
\def\bvec#1{\on\leftarrow{#1}}                  % backward vector accent
\def\oover#1{\on\circ{#1}}                              % circle accent

\def\[{\lfloor{\hskip 0.35pt}\!\!\!\lceil}
\def\]{\rfloor{\hskip 0.35pt}\!\!\!\rceil}
\def\Lag{{\cal L}}
\def\du#1#2{_{#1}{}^{#2}}
\def\ud#1#2{^{#1}{}_{#2}}
\def\dud#1#2#3{_{#1}{}^{#2}{}_{#3}}
\def\udu#1#2#3{^{#1}{}_{#2}{}^{#3}}
\def\calD{{\cal D}}
\def\calM{{\cal M}}

\def\szet{{${\scriptstyle \b}$}}
\def\ulA{{\un A}}
\def\ulM{{\underline M}}
\def\cdm{{\Sc D}_{--}}
\def\cdp{{\Sc D}_{++}}
\def\vTheta{\check\Theta}
\def\fracm#1#2{\hbox{\large{${\frac{{#1}}{{#2}}}$}}}
\def\ha{{\fracmm12}}
\def\tr{{\rm tr}}
\def\Tr{{\rm Tr}}
\def\itrema{$\ddot{\scriptstyle 1}$}
\def\ula{{\underline a}} \def\ulb{{\underline b}} \def\ulc{{\underline c}}
\def\uld{{\underline d}} \def\ule{{\underline e}} \def\ulf{{\underline f}}
\def\ulg{{\underline g}}
\def\items#1{\\ \item{[#1]}}
\def\ul{\underline}
\def\un{\underline}
\def\fracmm#1#2{{{#1}\over{#2}}}
\def\footnotew#1{\footnote{\hsize=6.5in {#1}}}
\def\low#1{{\raise -3pt\hbox{${\hskip 0.75pt}\!_{#1}$}}}

\def\Dot#1{\buildrel{_{_{\hskip 0.01in}\bullet}}\over{#1}}
\def\dt#1{\Dot{#1}}

\def\DDot#1{\buildrel{_{_{\hskip 0.01in}\bullet\bullet}}\over{#1}}
\def\ddt#1{\DDot{#1}}

\def\DDDot#1{\buildrel{_{_{\hskip 0.01in}\bullet\bullet\bullet}}\over{#1}}
\def\dddt#1{\DDDot{#1}}

\def\DDDDot#1{\buildrel{_{_{\hskip 
0.01in}\bullet\bullet\bullet\bullet}}\over{#1}}
\def\ddddt#1{\DDDDot{#1}}

\def\Tilde#1{{\widetilde{#1}}\hskip 0.015in}
\def\Hat#1{\widehat{#1}}

% Aligned equations

\newskip\humongous \humongous=0pt plus 1000pt minus 1000pt
\def\caja{\mathsurround=0pt}
\def\eqalign#1{\,\vcenter{\openup2\jot \caja
        \ialign{\strut \hfil$\displaystyle{##}$&$
        \displaystyle{{}##}$\hfil\crcr#1\crcr}}\,}
\newif\ifdtup
\def\panorama{\global\dtuptrue \openup2\jot \caja
        \everycr{\noalign{\ifdtup \global\dtupfalse
        \vskip-\lineskiplimit \vskip\normallineskiplimit
        \else \penalty\interdisplaylinepenalty \fi}}}
\def\li#1{\panorama \tabskip=\humongous                         % eqalignno
        \halign to\displaywidth{\hfil$\displaystyle{##}$
        \tabskip=0pt&$\displaystyle{{}##}$\hfil
        \tabskip=\humongous&\llap{$##$}\tabskip=0pt
        \crcr#1\crcr}}
\def\eqalignnotwo#1{\panorama \tabskip=\humongous
        \halign to\displaywidth{\hfil$\displaystyle{##}$
        \tabskip=0pt&$\displaystyle{{}##}$
        \tabskip=0pt&$\displaystyle{{}##}$\hfil
        \tabskip=\humongous&\llap{$##$}\tabskip=0pt
        \crcr#1\crcr}}

% Some defs

\def\eV{\,{\rm eV}}
\def\keV{\,{\rm keV}}
\def\MeV{\,{\rm MeV}}
\def\GeV{\,{\rm GeV}}
\def\TeV{\,{\rm TeV}}
\def\sv{\left<\sigma v\right>}
\def\({\left(}
\def\){\right)}
\def\cm{{\,\rm cm}}
\def\K{{\,\rm K}}
\def\kpc{{\,\rm kpc}}
\def\beq{\begin{equation}}
\def\eeq{\end{equation}}
\def\bea{\begin{eqnarray}}
\def\eea{\end{eqnarray}}

% New commands

\newcommand{\be}{\begin{equation}}
\newcommand{\ee}{\end{equation}}
\newcommand{\nbe}{\begin{equation*}}
\newcommand{\nee}{\end{equation*}}

\newcommand{\fr}{\frac}
\newcommand{\lb}{\label}

\thispagestyle{empty}

{\hbox to\hsize{
\vbox{\noindent May 2025 \hfill IPMU25-0001} }}

\noindent  

\noindent
\vskip2.0cm
\begin{center}

{\large\bf On Legacy of Starobinsky Inflation}
\vglue.2in

{\it Invited Contribution to the Starobinsky Memorial Volume, Springer 2025} 
\vglue.3in

Sergei V. Ketov~${}^{a,b,c,\#}$
\vglue.3in

${}^a$~Department of Physics, Tokyo Metropolitan University,\\
1-1 Minami-ohsawa, Hachioji-shi, Tokyo 192-0397, Japan \\
${}^b$~Department of Physics and Interdisciplinary Research Laboratory, \\Tomsk State University,
Tomsk 634050, Russian Federation\\
${}^c$~Kavli Institute for the Physics and Mathematics of the Universe (WPI),
\\The University of Tokyo Institutes for Advanced Study,  Kashiwa 277-8583, Japan\\

\vglue.2in

${}^{\#}$~ketov@tmu.ac.jp
\end{center}

\vglue.4in

\begin{center}
{\Large\bf Abstract}  
\end{center}
\vglue.1in

\noindent  Alexei Alexandrovich Starobinsky was one of the greatest cosmologists of all times, who made fundamental contributions to
gravitational theory and cosmology based on geometrical ideas in physics, in the spirit of Einstein. One of his big achievements is the famous Starobinsky model of cosmological inflation in the early universe, proposed in 1979-1980. In this memorial paper, the Starobinsky inflation model is systematically reviewed from the modern perspective. Its deformation to include production of primordial black holes is proposed, and possible quantum corrections in the context of superstring theory and the Swampland Program are discussed. Starobinsky inflation also leads to the universal reheating mechanism for particle production after inflation.

\newpage

\section{Introduction}

My research collaboration and personal encounters with Alexei Alexandrovich Starobinsky were both accidental and inevitable. In order to explain the apparent contradiction, I briefly recall where I came from. 

I was raised in a tiny remote village in Western Siberia, in the former Soviet Union. Despite that, it was possible to get there by mail the high-quality journals for school children, known as the "Quant "and the "Pioneer", where advanced articles with exercises (problems) in physics and mathematics were published. Some of those problems included the names of their inventors, and among  them were e.g., Ya.B. Zeldovich and A.N. Kolmogorov, just to mention a couple of famous scientists. There were regular competitions in physics and mathematics also, organized by university teachers. After winning some of those competitions, I was sent to a high school specialized in physics and mathematics and later to a university without passing  entrance exams because my teachers in the high school were university professors and I already completed the first-year university program in physics and mathematics, being in the high school. In the middle of my university studies, I was sent to the Physical Institute (FIAN) of the Soviet Academy of Sciences in Moscow, known as the Lebedev Physical Institute now. Being attached to the Theory Laboratory headed by V. L. Ginzburg, another famous scientist, and attending numerous seminars there, I was thinking about a research topic for my graduation. There were two natural options because the Ginzburg Lab had the cosmology group with prominent scientists such as A.D. Sakharov, A.D. Linde and V.F. Mukhanov, and the quantum field theory group including other prominent scientists such as E.S. Fradkin, M.A. Vasiliev, I.A. Batalin and I.V. Tyutin. I have chosen to conduct my research in supersymmetric quantum field theory and string theory under the supervision of E.S. Fradkin because this topic had a solid mathematical foundation and promising connections to high energy particle physics unlike cosmology that lacked a reliable theoretical and experimental foundation in the late 70's and the early 80's of the last century. Much later, around 2010, being in Tokyo, I turned to theoretical cosmology of the early universe because my research path in formal high-energy physics naturally guided me there. At  the same time, A.A. Starobinsky was a Visiting Professor at the University of Tokyo, and he approached me with a question about a realization of inflation in supergravity. Our research collaboration began with that question and resulted in joint papers proposing new supergravity-based realizations of inflation \cite{Ketov:2010qz,Ketov:2012jt}. Over the last 16 years my research group developed this subject by including spontaneous supersymmetry breaking, production of primordial black holes, dark energy, dark matter and gravitational waves 
\cite{Gates:2009hu,Ketov:2009wc,Ketov:2009sq,Ketov:2010eg,Ketov:2011rf,Ketov:2011tm,Ketov:2012yz,Ketov:2012se,Ketov:2012kf,Ketov:2013sfa,Ketov:2013dfa,Ketov:2014qoa,Ketov:2014qha,Ketov:2014hya,Addazi:2017rkc,Addazi:2017ulg,Aldabergenov:2017hvp,Addazi:2018pbg,Ketov:2018uel,Abe:2018rnu,Ketov:2019qjw,Ketov:2019mfc,Aldabergenov:2019aag,Ketov:2019rzg,Ketov:2019toi,Aldabergenov:2020pry,Ketov:2020hub,Aldabergenov:2020bpt,Aldabergenov:2020yok,Ketov:2021fww,Aldabergenov:2022rfc,Ketov:2022zhp,Ketov:2023ykf,Ishikawa:2024xjp,Frolovsky:2024xet}.

In this memorial paper, Starobinsky inflation is reviewed without supersymmetry and supergravity for simplicity. It is needed and actual because legacy of Starobinsky inflation is still not settled since the groundbreaking first paper \cite{Starobinsky:1980te}  appeared almost 45 years ago. 

In Section 2, the Starobinsky model of inflation, its possible quantum origin and its predictions for observations and measurements of the cosmic microwave background (CMB) radiation are reviewed from the modern perspective. In Section 3, a phenomenological deformation of the Starobinsky inflation model in the framework of modified $F(R)$ gravity is given in order to accommodate production of primordial black holes (PBH) during inflation.  In Section 4, the basic conjectures of Swampland Program in application to Starobinsky inflation are briefly discussed. An impact of the leading (closed) superstring correction to the low-energy gravitational effective action (EFT) on Starobinsky inflation is studied in Section 5. Section 6 is devoted to reheating after Starobinsky inflation.  Section 7 is a conclusion where
the future prospects are outlined.

\section{Starobinsky inflation model}

In this Section, the Starobinsky model of inflation is described from the current perspective (at present) without following the historical
developments that were rather confusing.

The Starobinsky model of inflation is the generally covariant and non-perturbative modified gravity theory whose Lagrangian is a sum of two terms, the Einstein-Hilbert (EH) term and the term quadratic in the Ricci scalar curvature $R$ with a positive coefficient. All the curvature-dependent terms beyond the EH one are irrelevant in the Solar system, while they may also be negligible during reheating after inflation in the weak-gravity regime. However, it is not the case during inflation in the high curvature regime where the $R^2$ term is the dominant contribution and the EH term can be ignored in the leading approximation.

Therefore, the Starobinsky model can be treated as the particular example of the modified $F(R)$ gravity theories that are geometrical because only gravitational interactions or General Relativity (GR) are used. A modified gravity action has the higher derivatives and generically suffers from Ostrogradsky instabilities and ghosts. However, in the most general modified gravity action, whose Lagrangian is quadratic in the spacetime curvature, the  {\it only} ghost-free term is just given by $R^2$ with a positive coefficient, which immediately leads to the Starobinsky model with the action 
\be  S_{\rm Star.} = \alpha \int \mathrm{d}^4x\sqrt{-g} R^2 + \fracmm{M^2_{\rm Pl}}{2}\int \mathrm{d}^4x\sqrt{-g} R~,\quad \alpha \equiv \fracmm{M^2_{\rm Pl}}{12M^2} ~, \lb{star}
\ee
with just one free parameter $\a$ or $M$, where $M_{\rm Pl}=1/\sqrt{8\p G_{\rm N}}\approx 2.4\times 10^{18}$ GeV, the spacetime signature is  $(-,+,+,+,)$ and the natural units are used, $\hbar=c=1$. The first term in this action is {\it scale invariant} because it does not have a dimensional coupling constant (the parameter $\a$ is dimensionless).

The origin of the $R^2$ term was initially proposed due to one-loop contributions of quantized matter fields in the EH gravity  \cite{Starobinsky:1980te}. However, because the EH term was sub-leading during inflation, the alternative interpretation is also possible, namely, with the EH term originated from the scale-invariant gravity with matter after quantization. For instance, when starting from 
the scale-invariant action for gravity and a scalar field $\phi$ as
\cite{Cooper:1981byv,Buchbinder:1986wk,Einhorn:2015lzy}
\be S[g_{\m\n},\phi] =  \int \mathrm{d}^4x\sqrt{-g} \left[ \alpha R^2 + \x \phi^2 R -\ha (\pa\phi)^2 -\l \phi^4\right]~, \lb{colw}
\ee
one finds that it can undergo a phase transition (called dimensional transmutation) due to quantum corrections, known as the 
 Coleman-Weinberg mechanism of spontaneous symmetry breaking \cite{Coleman:1973jx}. It leads to the massive scalar field $\phi$ that may  be identified with dilaton or Higgs field having  a non-vanishing vacuum expectation value (VEV)  in the effective action, as can be demonstrated in the one-loop perturbation theory \cite{Buchbinder:1986wk,Einhorn:2015lzy}. As a result, both the Planck mass and the EH term are generated with 
\be \ha M^2_{\rm Pl} = \x \VEV{\phi}^2~.
\ee
The alternative interpretation with the induced EH term is viable, though the action  (\ref{colw}) is still nonrenormalisable and, hence, cannot be considered as the UV-completion of the Starobinsky modified gravity.

The background metric of a flat Friedman (early) universe is given by 
\be ds^2=-dt^2+a^2\left(dx_1^2+dx_2^2+dx_3^2\right)~,\lb{flatF}
\ee
whose cosmic factor $a(t)$ is time-dependent. The action (\ref{star}) with the metric (\ref{flatF}) leads to equations of motion in the form
\be   \lb{stareom}
2H\ddot{H} - \left(\dot{H}\right)^2 + H^2\left(6\dot{H} + M^2\right)=0~,\quad H=\dot{a}/a~,
\ee
known as the Starobinsky equation in the literature, where the dots stand for the time derivatives and $H(t)$ is Hubble function.

When searching for a solution to the Starobinsky equation in the form of left Painlev\'e series, 
$H(t)=\sum^{k=p}_{k=-\infty}c_k(t_0-t)^k$, one finds the Hubble function (see e.g., Ref.~\cite{Ketov:2022zhp})
\be \lb{stars}
\begin{split}
\fracmm{H(t)}{M}  & =  \fracmm{M}{6}(t_0-t)+\fracmm{1}{6M(t_0-t)} - \fracmm{4}{9M^3(t_0-t)^3}+
\fracmm{146}{45M^5(t_0-t)^5}  \\
& {} -\fracmm{11752}{315 M^7 (t_0-t)^7} + {\cal O} \left(M^{-9}(t_0-t)^{-9}\right)
\end{split}
\ee
valid for $M(t_0-t)>1$. This special solution is an attractor, while $R=12H^2+6\dot{H}$.

Therefore, in the slow-roll (SR) approximation defined by $\abs{\ddot{H}} \ll \abs{H\dot{H}}$ and  $\abs{\dot{H}} \ll H^2$, the leading term in the Starobinsky solution reads
\be 
 H(t) \approx \left (\fracmm{M^2}{6}\right) (t_0-t)~,
\ee
being entirely due to the $R^2$-term in the action. The attractor solution spontaneously breaks the scale invariance of the $R^2$-gravity and, therefore, implies the existence of the Nambu-Goldstone boson (called scalaron) that is the physical excitation of the higher-derivative Starobinsky gravity. It can be made manifest by rewriting the Starobinsky action to the quintessence form after the field redefinition (or the Legendre-Weyl transform) with \cite{Maeda:1988ab}
\begin{equation} 
  \varphi =  \sqrt{ \fracmm{3}{2}} M_{\rm Pl}\ln F'(\c)   \quad {\rm and}\quad g_{\m\n}\to \fracmm{2}{M^2_{\rm Pl}}F'(\chi) g_{\m\n}~,
  \quad \chi=R~.
  \ee
It yields the standard quintessence action
\be S[g_{\m\n},\varphi]  = \fracmm{M^2_{\rm Pl}}{2}\int \mathrm{d}^4x\sqrt{-g} R 
 - \int \mathrm{d}^4x \sqrt{-g} \left[ \frac{1}{2}g^{\m\n}\pa_{\m}\varphi\pa_{\n}\varphi
 + V(\varphi)\right] \lb{quint}
\ee
in terms of the canonical inflaton $\varphi$ with the scalar potential 
\be \lb{starp}
V(\varphi) = \fracmm{3}{4} M^2_{\rm Pl}M^2\left[ 1- \exp\left(-\sqrt{\frac{2}{3}}\varphi/M_{\rm Pl}\right)\right]^2~.
\ee
This potential has the infinite plateau (for the large $\varphi$-field values of the order $M_{\rm Pl}$ and beyond)  with the approximate shift symmetry of the inflaton field as the consequence of the scale invariance of the $R^2$ gravity or the approximate scale invariance of the action (\ref{star}) in the large-curvature regime. The potential (\ref{starp}) also has the positive "cosmological constant" given by the first term in the square brackets, induced by the $R^2$ term in the action (\ref{star}), which can be physically interpreted as the energy driving inflation. The scale of inflation is determined by the mass parameter $M$ that is identified with the canonical inflaton mass. The universality class of the Starobinsky-like inflation models is determined by the critical parameter $\sqrt{2/3}$ in the exponential term \cite{Ketov:2021fww}.

The equivalent actions (\ref{star}) and (\ref{quint}) are usually referred to the Jordan frame and the Einstein frame, respectively, in the literature. The approximate shift symmetry of the potential (\ref{starp}) required for proper duration of inflation arises as the consequence of the approximate scale invariance of the Starobinsky gravity. The $R^2$-term is unique among {\it all} the higher-derivative curvature terms because it is the only one that leads to a ghost-free scale-invariant action. Therefore, the $R^2$ term must be present in any viable model of inflation based on modified gravity. This conclusion becomes more transparent by using the {\it inverse} transformation from  the quintessence action (with a potential $V$) in the Einstein frame to an $F(R)$ gravity action in the Jordan frame, having the parametric form \cite{Ketov:2014jta}
\be
R = \left(  \fracmm{\sqrt{6}}{M_{\rm Pl}}
    \fracmm{d V}{d \varphi} + \fracmm{4V}{M^2_{\rm Pl}} \right) e^{ \sqrt{\frac{2}{3}} 
  \varphi/M_{\rm Pl}}~,    \quad F= \left(  \fracmm{\sqrt{6}}{M_{\rm Pl}}
 \fracmm{d V}{d \varphi} + \fracmm{2V}{M^2_{\rm Pl}} \right) e^{ 2 \sqrt{\frac{2}{3}} \varphi/M_{\rm Pl}}~,
\ee
with the inflaton $\varphi$ as the parameter. In the SR approximation, the first term in the brackets is negligible against the second term, which immediately implies $F(R)\sim R^2$. 

The gravitational EFT during inflation does not have to be limited to the terms given in Eq.~(\ref{star})  but can also include 
the higher-order curvature terms. Those terms eliminate the infinite plateau in the inflaton potential (\ref{starp}) and may destabilize
Starobinsky inflation. However, the fact that the Starobinsky model of inflation is in excellent agreement with the  current CMB measurements (see below) implies those terms should also be sub-leading during Starobinsky inflation, which put restrictions on their (unknown) coefficients.

It is convenient to use the e-folds number $N$ instead of time $t$, which are related by
\be 
N=\int^{t_0}_{t} H(\tilde{t}) d\tilde{t}~,
\ee
and the co-moving wavenumber $k=2\pi/\lambda$ as a scale.  The scale $k$ is related to e-folds $N$ by the equation $d\ln k =-dN$~.

The SR (running) parameters in the Einstein frame are defined by
\be
\ve_{\rm sr}(\varphi) = \fracmm{M^2_{\rm Pl}}{2}\left( \fracmm{V'}{V}\right)^2 \quad {\rm and} \quad 
\eta_{\rm sr}(\varphi) = M^2_{\rm Pl} \left( \fracmm{V''}{V}\right)
\ee
in terms of the quintessence scalar potential $V$, where the primes denote the derivatives with respect to canonical inflaton $\varphi$. In  the Jordan frame, one uses the Hubble flow functions,
\be \lb{Hflow}
\epsilon_{H} = -\fracmm{\dot{H}}{H^{2}}~,\quad 
	\eta_{H} = \epsilon_{H} - \fracmm{\dot{\epsilon}_{H}}{2\epsilon_{H} H}~~.
\ee

The amplitude of scalar perturbations at the horizon crossing with the pivot scale $k_*=0.05~{\rm Mpc}^{-1}$  is known from CMB
measurements (called WMAP or COBE normalization) as
\be 
A_s= \fracmm{V_*^3}{12\p^2 M^6_{\rm Pl}({V_*}')^2}=\fracmm{3M^2}{8\p^2M^2_{\rm Pl}}\sinh^4\left(
\fracmm{\varphi_*}{\sqrt{6}M_{\rm Pl}}\right)\approx 2 \cdot 10^{-9}~,
\ee
where subscript (*) refers to the CMB pivot scale in the case of Starobinsky inflation. Therefore, the known amplitude $A_s$ determines the parameter $M$ (or $\a$) as well as the scale of inflation, $H_{\rm inf.}$, in the Starobinsky model as 
\be
\fracmm{M}{M_{\rm Pl}}\approx   {\cal O}(10^{-5}), \quad \a\approx {\cal O}(10^9), \quad H\approx {\cal O}(10^{14})~{\rm GeV},\quad 
\fracmm{R}{M^2_{\rm Pl}} \approx \fracmm{12H^2}{M^2_{\rm Pl}}\approx 10^{-7}.
\ee

The higher-order curvature terms in the gravitational EFT beyond the Starobinsky model of inflation lead to power series in $H^2/M^2_{\rm Pl}\sim 10^{-8}$. Those terms are sub-leading during inflation unless they have very large coefficients. It is also worth noticing that the large value of $\alpha$ required by CMB does not speak in favor of generating 
the $R^2$-term by quantum matter contributions because a single quantized matter field contributes in the one-loop approximation about
$10^{-3}$ to the $\alpha$-parameter, so one needs about $10^{12}$ quantized matter fields in order to achieve the value of $\a$ needed.

The primordial spectrum $P_{\z}(k)$ of 3-dimensional scalar (density) perturbations $\z(x)$ in a flat Friedman universe is defined by the 2-point correlation function 
\be \lb{defpsp}
\VEV{ \fracmm{\d\z(x)}{\z} \fracmm{\d\z(y)}{\z}} =\int\fracmm{d^3k}{k^3} e^{ik\cdot (x-y)}\fracmm{P_{\z}(k)}{P_0}~~,
\ee
where $k$ is the comoving wavenumber (or scale). Similarly, one defines the primordial spectrum $P_t(k)$ of tensor perturbations, 
see e.g., Ref.~\cite{Ketov:2021fww} for details. The power spectra define the amplitudes $A_s(k)$ and $A_t(k)$, respectively. 

Given the power spectra $P_{\z}(k)$ and $P_t(k)$,  one defines the scalar tilt $n_s(k)$, its running parameter $\a_s(k)$, the tensor tilt $n_t(k)$ and its running parameter $\a_t(k)$ (all dimensionless) as
\be \lb{tilts}
n_s = 1 + \fracmm{d\ln P_{\z}(k)}{d\ln k}~, \quad \a_s =  \fracmm{d^2\ln P_{\z}(k)}{(d\ln k)^2}~,\quad n_t = \fracmm{d\ln P_{t}(k)}{d\ln k}~, 
 \quad \a_t =  \fracmm{d^2\ln P_t(k)}{(d\ln k)^2}~~,
 \ee
 as well as the tensor-to-scalar ratio 
\be \lb{ts}   r(k) = \fracmm{P_t}{P_\z}=8\abs{n_t}~.
\ee

The Starobinsky model  gives sharp predictions for the cosmological tilts of the CMB scalar and tensor power spectra.  {\it In the leading order} with respect to the inverse e-folds $N_*$ evaluated when perturbations left the horizon (at the horizon crossing), those predictions are \cite{Mukhanov:1981xt,Mukhanov:1990me}
\be 
n_s\approx 1- \fracmm{2}{N_*}~~,\quad \a_s\approx -\fracmm{2}{N^2_*}~~,\quad \a_t\approx -\fracmm{3}{N^3_*}~,\quad 
r\approx \fracmm{12}{N^2_*}~~. \lb{strtilts}
\ee
Therefore, tensor perturbations are suppressed with respect to scalar perturbations by the extra factor of $N^{-1}_*$. The theoretical predictions (\ref{strtilts}) are to be compared to the recent (Planck mission) CMB measurements \cite{BICEP:2021xfz,Tristram:2021tvh},
which give 
\be  \lb{cmb}
n_s\approx 0.9649\pm 0.0042~(68\% {\rm CL}) \quad {\rm and} \quad r< 0.032~(95\% {\rm CL})~,
\ee
and fit the Starobinsky model predictions for 
\be \lb{Nstar}
N_*=56 \pm 8~.
\ee
This prediction for the duration of inflation agrees with the calculations in Ref.~\cite{Toyama:2024ugg} in the Jordan frame, based on the solution (\ref{stars}). The corresponding times for the end and the beginning of inflation are $M(t_0-t_{\rm end})\approx 2.5$ and $M(t_0-t_{\rm start})\approx 27.7$, respectively. 

It is difficult to independently predict or measure the duration of inflation. However, excluding $N_*$ from Eqs.~(\ref{strtilts}) yields the 
prediction of the Starobinsky model  for the tensor-to-scalar ratio with high precision as
\be r\approx 3(1-n_s)^2~. \lb{starrel}
\ee

The Starobinsky model is sensitive to quantum ultra-violet (UV) corrections because of its high inflation scale and the inflaton field values near the Planck scale during inflation. Therefore, it is important to determine the UV-cutoff $\L_{\rm UV}$ of the Starobinsky model by studying scaling of inflaton scattering amplitudes with respect to energy, $E/\L_{\rm UV}$.
A careful calculation yields \cite{Hertzberg:2010dc}
\be \lb{uvcut}
\Lambda_{\rm UV}=M_{\rm Pl}~.
\ee
Therefore, the predictions of the Starobinsky model for inflation and CMB make sense and the model itself can be considered as a trustable EFT after decoupling of heavy modes expected at the Planck scale \cite{Ketov:2024klm}.

It is worth mentioning that very little from this Section can be found in the original paper \cite{Starobinsky:1980te} that is the
standard reference for Starobinsky inflation. The key point of Ref.~\cite{Starobinsky:1980te} was a discovery that quantum-modified gravitational equations of motion can have attractor solutions suitable for describing inflation.

\section{Deformation of the Starobinsky inflation model for production of primordial black holes at smaller scales}

The Starobinsky model can be extended within modified $F(R)$ gravity in order to describe double slow-roll (SR) inflation with an ultra-slow-roll (USR) phase by engineering the function $F(R)$ leading to a near-inflection point in the inflaton potential below the scale of inflation \cite{Frolovsky:2022ewg,Saburov:2023buy}. It results in large density perturbations whose gravitational collapse leads to production  of primordial black holes (PBH). However, adding the near-inflection point and the USR phase requires fine-tuning of the model parameters \cite{Geller:2022nkr}, which often lowers the value of the CMB tilt $n_s$ of scalar perturbations and thus leads to tension with CMB measurements, while large perturbations may also imply significant nonGaussianity and large quantum (loop) corrections that may invalidate classical single-field models of inflation with PBH production \cite{Kristiano:2022maq}. 

The phenomenological model \cite{Frolovsky:2022ewg,Saburov:2023buy} is an extension of the singularity-free Appleby-Battye-Starobinsky (ABS) model \cite{Appleby:2009uf}  with other parameters whose fine-tuning leads to efficient production of PBH  having asteroid-size masses (and atomic sizes) between $10^{16}$ g and $10^{20}$ g, exceeding the Hawking (black hole) evaporation limit of $10^{15}$ g. Those PBH may be part (or the whole) of the current dark matter (DM), while the model itself may be confirmed or falsified by detection or absence of the induced gravitational waves with frequencies in the $10^{-2}$ Hz range (see below).

The $F(R)$ gravity model \cite{Saburov:2023buy} is defined by the action $(M_{\rm Pl}=1$) 
\be \lb{MGaction}
 S = \fracm{1}{2} \int d^4x \sqrt{-g} \,F(R)~~,
\ee
whose $F$-function of the spacetime scalar curvature $R$ reads
\begin{equation} \lb{Ff}
	F(R)=(1+g\tanh b) R + gE_{AB}  \ln \left[\fracmm{\cosh \left(\frac{R}{E_{AB}}-b\right)}{\cosh (b)}\right]
	+\fracmm{R^2}{6 M^2} - \d\fracmm{R^4}{48M^6}~~,
\end{equation}
where the first terms are known in the literature as the ABS model \cite{Appleby:2009uf} with 
the Starobinsky mass $M\approx 1.3\times 10^{-5}$ defining the scale of the first SR phase of inflation.The ABS parameter 
\begin{equation} \lb{ABp}
	E_{AB}=\fracmm{R_0}{2g\ln(1+e^{2b})}
\end{equation}
has the scale $R_0$ related to the second SR phase of inflation below the Starobinsky scale, while it differs from Ref.~\cite{Appleby:2009uf} where $R_0$ was associated with the dark energy scale.  The other parameters $g$ and $b$ define the shape of the inflaton potential and have to be fine-tuned in order to engineer a near-inflection point. The last term in Eq.~(\ref{Ff}) may be considered as a quantum gravity correction (see  Section 6) in order to get good (within $1\sigma$) agreement  with the measured CMB value of $n_s$. The function (\ref{Ff}) obeys the no-ghost (stability) conditions, $F'(R)>0$ and $F''(R) >0$, for the relevant values of $R$, avoids singularities, and has the correct Newtonian limit.

To produce PBH, one needs a large enhancement of the power spectrum of scalar perturbations by seven orders of magnitude against the CMB spectrum. Then the parameters $(R_0,g,b)$ should be fine-tuned as \cite{Saburov:2023buy}
\be \lb{parv}
R_0\approx 3.00M^2~,\quad g\approx 2.25 \quad {\rm and} \quad b\approx 2.89~.
\ee
The numerically obtained profile of the inflaton potential $V(\f)$ (in the Einstein frame) for some values of $R_0$ and $\d$ is given in Fig.~\ref{ris:V1}.

\begin{figure}[h]
\begin{minipage}[h]{0.5\linewidth}
\center{\includegraphics[width=0.8\linewidth]{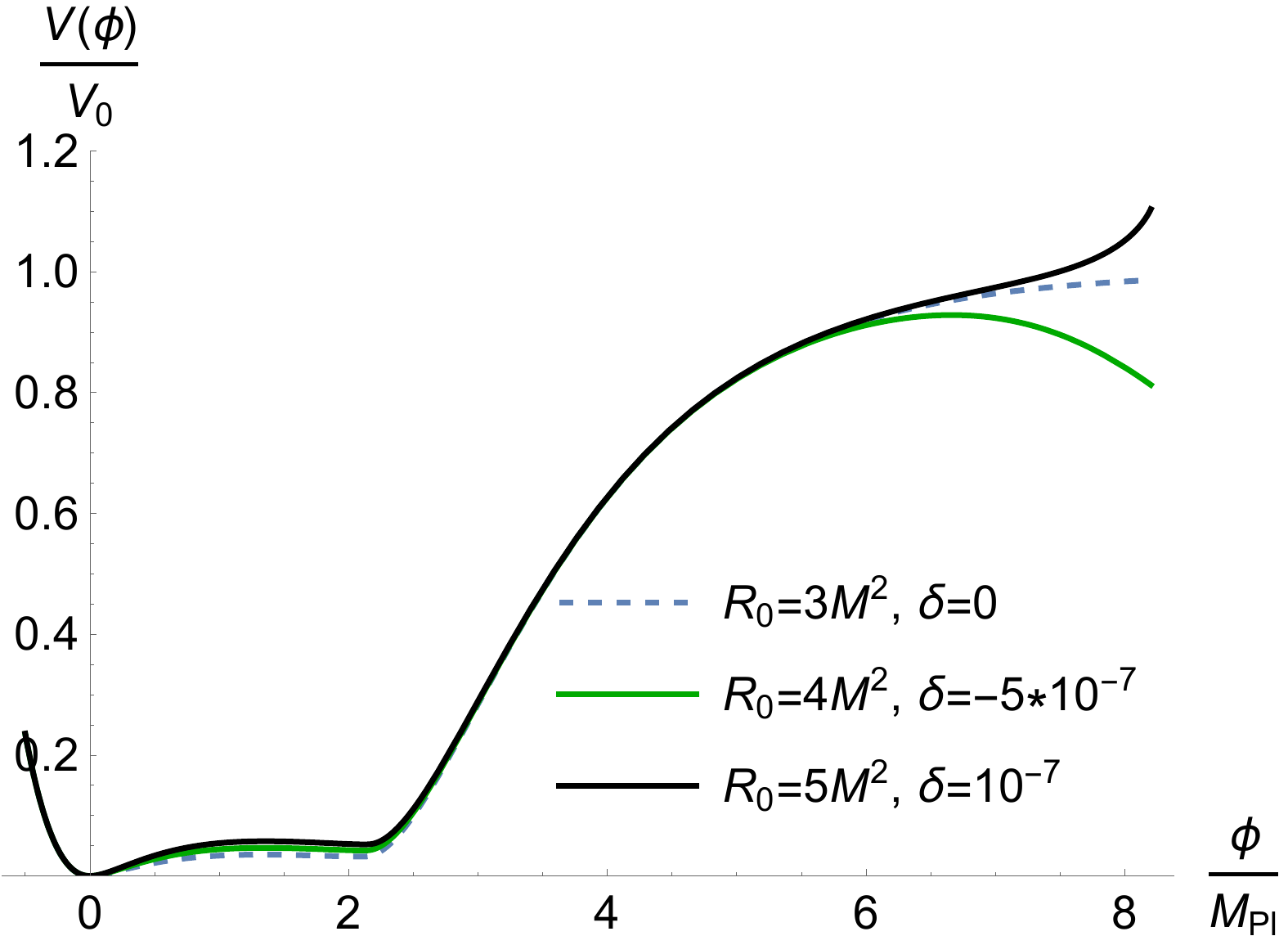} }
\end{minipage}
\hfill
\begin{minipage}[h]{0.5\linewidth}
\center{\includegraphics[width=0.8\linewidth]{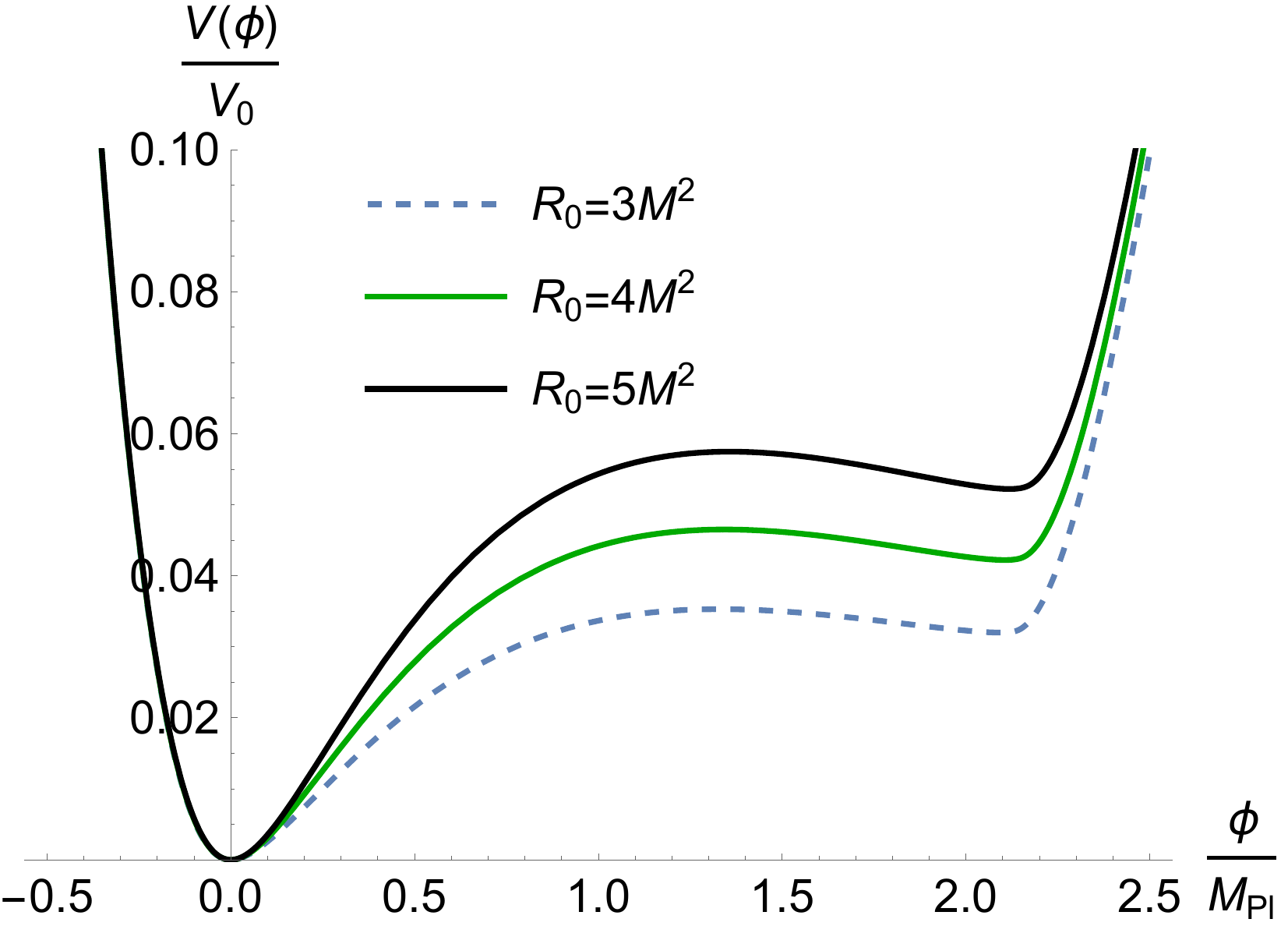} }
\end{minipage}
\caption{The inflaton potential having two plateaus for $g=2.25$ and $b=2.89$ with $V_0=\fracmm{3}{4}M^2$ (left). Zooming the potential for lower values of $\f$ with a near-inflection point (right). The potential is unstable for negative values of $\d$, and has the infinite plateau for $\d=0$ leading to Starobinsky inflation.}
\label{ris:V1}
\end{figure}

The CMB observables with the model best fit are given by~\cite{Saburov:2023buy}
\be \lb{keyobs}
n_s\approx 0.965, \quad r\approx 0.0095, ~\quad M_{\rm PBH}\approx 1.0 \cdot 10^{20}~{\rm g}.
\ee

The PBH-producing peak in the scalar power-spectrum of Fig.~\ref{ris:P0} can be approximated by the log-normal fit \cite{Pi:2020otn,Frolovsky:2023xid} as
\be \lb{lognorps}
P^{\rm peak}_{\z}(k) \approx \fracmm{A_{\z}}{\sqrt{2\p} \D} \exp \left[ \fracmm{-\ln^2(k/k_p)}{2\D^2}\right]
\ee
with the amplitude $A_{\z}\approx 0.06$  and the width $\D\approx 1.5$, where $k_p\approx 4.5\cdot 10^{12}~{\rm Mpc}^{-1}$ is the location of the peak.

\begin{figure}[h]
\center{\includegraphics[width=0.5\linewidth]{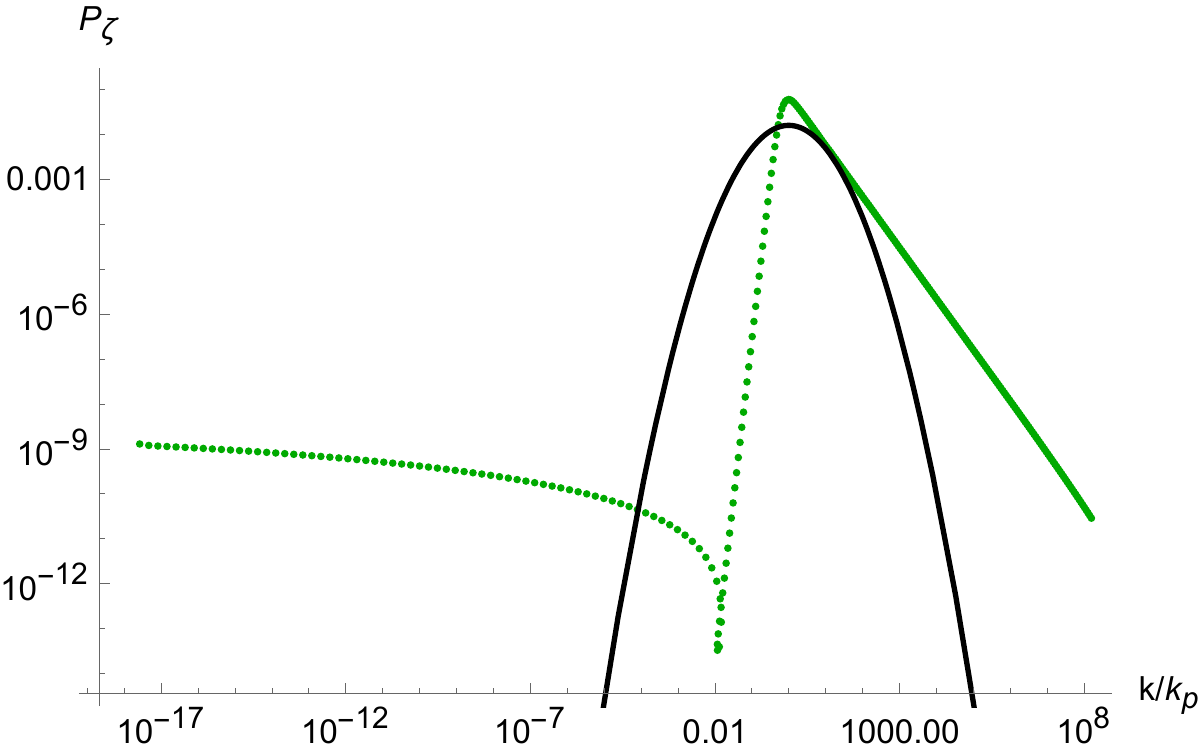}}
\caption{The primordial power spectrum $P_{\z}(k)$ of scalar perturbations and its log-normal fit.}
\label{ris:P0}	
\end{figure}

PBH production in the early universe leads to stochastic gravitational waves (GW) different from primordial GW caused by inflation. 
The PBH-induced GW density in the model also can be approximated by the lognormal fit as \cite{Pi:2020otn,Saburov:2024und}
\be \lb{logngw}
  \O_{\rm GW}(k)=\fracmm{A_{\rm GW}}{\sqrt{2\pi}\sigma_{\rm GW}}\mathrm{exp}\left[-\fracmm{\mathrm{ln}^2(k/k_p)}{2\sigma^2_{\rm GW}}\right]~,
\ee
with the amplitude $A_{\rm GW}\approx 5.6\cdot 10^{-8}$ and the width $\sigma_{\rm GW}\approx \D/ \sqrt{2}\approx 1.06$,
where $\D$ is the width of the power spectrum in Fig.~\ref{ris:P0}. The $\Omega_{\mathrm{GW}}^{\rm peak}(k)$ near the peak is roughly given by $10^{-6}\mathcal{P}^2_\zeta(k)$.

The induced GW frequencies $f_p$ are related to the PBH masses as \cite{DeLuca:2020agl}
\be \lb{gwf}
  f_p\approx5.7\left(\fracmm{M_\odot}{M_{\mathrm{PBH}}}\right)^{1/2}10^{-9}~\mathrm{Hz},
\ee
where the Sun mass is given by $M_\odot\approx 2\cdot 10^{33}$ g. Given the PBH masses of $10^{20}$ g, it results in the GW frequency $f_p\approx 0.0255$~Hz that is higher than the GW frequencies between 3 and 400 nHz  detected by NANOGrav \cite{NANOGrav:2023gor}.  A more specific comparison of the model predictions with future GW observations is possible by plotting the GW spectrum in the model against the expected sensitivity curves in the future space-based gravitational interferometers such as LISA \cite{LISA:2017pwj}, TianQin \cite{TianQin:2015yph}, Taiji \cite{Gong:2014mca} and DECIGO \cite{Kudoh:2005as}, see Fig.~\ref{sensi3}. The  space-based experiments are expected to be sensitive to stochastic GW in the frequencies between $10^{-3}$ and $10^{-1}$ Hz, while  part of the black curve in Fig.~\ref{sensi3} from the model belongs to that frequency range also.

\begin{figure}[h]
  \centering
  \begin{minipage}[t]{0.45\hsize}
    \includegraphics[keepaspectratio, scale=1.0]{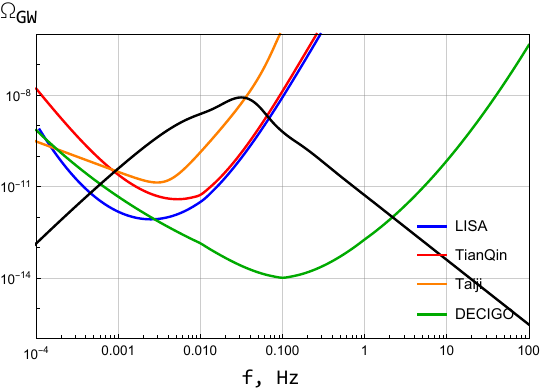}
  \end{minipage}
  \caption{The GW density induced by the power spectrum of scalar perturbations in the model (in black) against the expected sensitivity curves for the future space-based GW experiments (in color).}
\lb{sensi3}
\end{figure}

It was found in Ref.~\cite{Saburov:2024und} by using the $\delta N$ formalism that the relative size of quantum (loop) corrections to the power spectrum of scalar perturbations in the model is of the order $10^{-3}$ or less, so that the model of inflation and PBH production considered in this Section is not ruled out by the quantum loop corrections.

\section{Starobinsky model and Swampland conjectures}

The Starobinsky model is a {\it large-field} model whose inflaton field values are near the Planck scale during inflation. Large-field models of inflation are sensitive to quantum gravity that may ruin their consistency. Unfortunately, little is known about quantum gravity,
not even its energy scale. Should the quantum gravity scale be the same as the Planck scale, then Eq.~(\ref{uvcut}) implies robustness of 
Starobinsky inflation against quantum gravity corrections. However, should the quantum gravity scale be close to the Starobinsky inflation scale of $10^{14}$ GeV, then the predictions of Starobinsky inflation would not be reliable and the Starobinsky model could not be EFT, i.e. it would belong to Swampland \cite{Lust:2023zql}.

A string theory derivation of the Starobinsky inflation is unknown, though it is possible to get the effective scalar potential close to that in 
Eq.~(\ref{starp}) in the Einstein frame, see e.g., \cite{Blumenhagen:2015qda,Brinkmann:2023eph}.  

In the absence of full theory of quantum gravity, one can employ a set of its expected features, which is known as the Swampland Program \cite{Palti:2019pca,vanBeest:2021lhn}. The Swampland conjectures are all about consistency with quantum gravity (or UV completion) either in string theory or in general. The Swampland Program is aimed to discriminate between possible EFT.  

To the end of this  Section, the simplest Swampland conjectures are briefly discussed in relation to the Starobinsky model of inflation.
\vglue.1in
{\it No global symmetries in quantum gravity:}\\
This Swampland conjecture is about the absence of global (or rigid) symmetries in any fundamental theory of quantum gravity.  It claims that only local (or gauge) symmetries are possible in quantum gravity. This conjecture is fully in line with fundamental principles of
 General Relativity but may appear unusual for particle physics based on the Standard Model near or below the electro-weak scale.
 
 As regards the Starobinsky model, it has only approximate (global) scale invariance related to the $R^2$ term alone, which is obviously violated by  the EH term and any other possible higher-order term with respect to the spacetime curvature in the gravitational EFT. Therefore, the Starobinsky model does not violate the no-global-symmetry conjecture.
\vglue.1in

{\it Weak-gravity conjecture:}\\
In its simplest version, this conjecture claims that gravity is always the weakest force, for example, in comparison to electromagnetic force.
At first sight, Starobinsky inflation contradicts this conjecture because the Starobinsky inflation relies on the gravitational
force as the main force driving inflation in the Jordan frame. However, in the Einstein frame, Starobinsky inflation is mainly due to the inflaton selfinteraction described by the scalar potential (\ref{starp}), i.e. due to the inflaton force (or inflaton exchange). Therefore, there is no contradiction between the Starobinsky model of inflation and the weak-gravity conjecture.

One may also argue that during inflation there were no sources of electromagnetic force because there were no charged particles yet (they appeared during reheating after inflation).
\vglue.1in

{\it No-de-Sitter conjecture:}\\
claims that no de Sitter (dS) spacetime and no eternal inflation are possible in quantum gravity.
Again, at first sight, it is in contradiction to Starobinsky inflation because the potential (\ref{starp}) has the infinite plateau.
However, the infinite plateau in the Starobinsky model is destabilized by the higher-order terms with respect to the spacetime
curvature, which are certainly present in any UV-completion of the Starobinsky action (\ref{star}). A stronger version of the conjecture, excluding all locally flat or single-field inflaton potentials \cite{Obied:2018sgi}, is apparently in conflict with observations \cite{Kehagias:2018uem,Denef:2018etk,Kinney:2018nny}.
\vglue.1in
{\it Distance conjecture:} \\
about possible EFT from string theory implies constraints on the inflaton range $\D \vf$ during inflation and, hence, on the CMB tensor-to-scalar-ratio $r$, via the upper bound \cite{Scalisi:2018eaz}
\be 
\abs{\D \vf} \leq M_{\rm Pl}\ln \left( \fracmm{M_{\rm Pl}}{H_{\rm inf}}\right)  \approx \fracmm{M_{\rm Pl}}{2}\ln \left(\fracmm{2}{\p^2A_sr}\right)
\ee
that is complementary to the Lyth lower bound \cite{Lyth:1996im}
\be 
\abs{\D \vf} \geq {\cal O}(1)  \left( \fracmm{r}{0.01}\right)^{1/2} M_{\rm Pl}~.
\ee
Both constraints are in agreement with Starobinsky inflation that has $\abs{\D \vf}\approx 5.5 M_{\rm Pl}$ but severely constrain it.

\section{Quantum correction from superstring theory}

Superstring theory is a good candidate for the theory of quantum gravity, so it is natural to seek an UV completion of the Starobinsky model there. However, it turns out to be difficult because superstring theory is defined in 10 spacetime dimensions, whereas the Starobinsky model is defined in 4 space-time dimensions. Therefore, 6 spatial dimensions in superstring theory have to be hidden or compact, while the gravitational low-energy EFT in string theory is subject to large ambiguities related to spacetime metric redefinitions and compactification, see Refs.~\cite{Ketov:2017aau,Nakada:2017uka}  for the alternative proposals to hidden dimensions.

In the case of closed (type II) superstrings, the leading $(\alpha')^3$-correction beyond the EH term is given by the terms {\it quartic} in the space-time curvature, which were first derived by Grisaru and Zanon in 1985 from the vanishing four-loop renormalization group beta-function of the supersymmetric non-linear sigma model in two dimensions, describing propagation of a test superstring in the gravitational background \cite{Grisaru:1986vi}. The same quartic curvature terms arise from M-theory \cite{Green:1997as,Blumenhagen:2024ydy} after dimensional reduction. The dependence of the gravitational EFT in string theory upon Ricci scalar curvature (and Ricci tensor also) is known to be ambiguous, see e.g., Ref.~\cite{Ketov:2000dy}, because closed strings (generating gravity) are well defined only in Ricci-flat spacetimes by string scattering amplitudes (S-matrix).

By combining the Starobinsky model with the GZ correction one gets the Starobinsky-Grisaru-Zanon (SGZ) gravity defined by the action
\begin{equation} \lb{sgz}
    S_{\rm SGZ}[g]=\fracmm{M_{\text{Pl}}^2}{2}\int d^4x \sqrt{-g}\,\left(R+ \fracmm{1}{6M^2}R^2 -\fracmm{72\g}{M^6}Z\right), 
\end{equation}
where the GZ superstring correction reads \cite{Grisaru:1986vi} 
\begin{equation}
    72Z=\left(R^{\mu\rho\sigma\nu}R_{\lambda\rho\sigma\tau}+\frac{1}{2}R^{\mu\nu\rho\sigma}R_{\lambda\tau\rho\sigma}\right)R_{\mu}^{\,\,\,\alpha\beta\lambda}R^{\tau}_{\,\,\,\alpha\beta\nu}
\end{equation}
with the effective dimensionless (string) coupling constant $\g>0$. The value of $\gamma$ cannot be calculated from string theory because it depends upon a compactification from ten to four spacetime dimensions and unknown vacuum expectation value of string dilaton.

Due to the origin of SGZ gravity, the first two terms in the action (\ref{sgz}) should be considered nonperturbatively but the last (quartic) curvature terms should be treated as a perturbation, in the first order with respect to the coupling constant $\gamma$. In particular, no Ostrogradski ghosts arise. 

In a flat Friedman universe (\ref{flatF}) one finds \cite{Toyama:2024ugg}
\be Z= H^8  + 2H^6\dot{H} +\fracm{11}{6} H^4\dot{H}^2 +\fracm{2}{3}H^2\dot{H}^3 + \fracm{1}{12} \dot{H}^4~,
\ee
and the SGZ gravity equations of motion are
\be \lb{eom}
\eqalign{
& m^{6} H^{2} + 6m^{4} H^{2} \dot{H} + 2m^{4} H \ddot{H} -m^{4} \dot{H}^{2} \cr
& - 12 \gamma H^{8} + 132 \gamma H^{6} \dot{H} + 44 \gamma H^{5} \ddot{H} + 138 \gamma H^{4} \dot{H}^{2} \cr
& + 48 \gamma H^{3} \dot{H} \ddot{H} + 28 \gamma H^{2} \dot{H}^{3} + 12 \gamma H \dot{H}^{2} \ddot{H} - 3 \gamma \dot{H}^{4}=0~,\cr}
\ee
which extend the Starobinsky equation (\ref{stareom}) by the $\gamma$-dependent terms. A solution to this equation in the first order
with respect to the $\gamma$-parameter reads \cite{Toyama:2024ugg}
\be H(t) = H_0(t) +\g H_1(t)~,
\ee
where
\be \eqalign{
\fracmm{H_1(t)}{M}= & -\fracmm{1}{163296}M^7(t_0-t)^7-\fracmm{2}{2835}M^5(t_0-t)^5-\fracmm{391}{90720}M^3(t_0-t)^3-\fracmm{9061}{306180}M(t_0-t) \cr
& -\fracmm{127}{7776M(t_0-t)}-\fracmm{1931203}{5358150M^3(t_0-t)^3}+{\cal O}(M^{-5}(t_0-t)^{-5})~. \cr} \lb{gcor}
\ee
Accordingly, the Ricci scalar is given by \cite{Toyama:2024ugg}
\be  \eqalign{
\fracmm{R}{M^2}= & ~~\fracmm{M^2 (t_0-t)^2}{3}-\fracmm{1}{3}-\fracmm{4}{9 M^2(t_0-t)^2}+\fracmm{16}{5 M^4 (t_0-t)^4}-\fracmm{6908}{189 M^6 (t_0-t)^6} \cr
&  +\gamma \left[-\fracmm{M^8 (t_0-t)^8}{40824}-\fracmm{151 M^6 (t_0-t)^6}{58320}+\fracmm{143 M^4 (t_0-t)^4}{122472}-\fracmm{4163 M^2 (t_0-t)^2}{81648} \right. \cr
&  -\fracmm{68713}{7144200} -\fracmm{5109281}{5143824 M^2(t_0-t)^2}+\fracmm{17584432631}{964467000 M^4 (t_0-t)^4} \cr
& \left.  -\fracmm{75802186291}{300056400 M^6 (t_0-t)^6}+{\cal O}(M^{-8}(t_0-t)^{-8})\right]~.\cr}
\ee
It was found in Ref.~\cite{Toyama:2024ugg} that demanding the absence of ghosts and negative energy fluxes implies the upper bound
\be \lb{bound2}
\gamma\leq 1.12\times10^{-6}~.
\ee

It is instructive to compare contributions of the GZ quantum correction to the CMB observables against the classical
contributions beyond the leading order given by Eq.~(\ref{strtilts}) within the possible range of $N_*$ in Eq.~(\ref{Nstar}).
The subleading terms for the CMB observables in the Starobinsky inflation model are given by \cite{Bianchi:2024qyp}
\be \lb{subs}
\eqalign{
n_s = & ~~1- \fracmm{2}{N_*} + \fracmm{2.4}{N^2_*} - \fracmm{\ln(2N_*)}{6N^2_*} + {\cal O}\left( \fracmm{\ln 2N_*}{N^3_*}\right)~,\cr
\a_s = &-\fracmm{2}{N^2_*} +  {\cal O}\left( \fracmm{\ln 2N_*}{N^3_*}\right)~,\cr
r = & ~~\fracmm{12}{N^2_*} + \fracmm{2\ln (2N_*)}{N^3_*} - \fracmm{56.76}{N^3_*} + {\cal O}\left( N^{-4}_*\right)~,\cr
\a_t = & -\fracmm{3}{N^3_*}  +  {\cal O}\left( N^{-4}_*\right)~,\cr}
\ee
where $\ln 2N_*/N^3_* <  4.1\cdot 10^{-5}$ and $N^{-4}_* < 2\cdot 10^{-7}$. The subleading contribution to the scalar tilt $n_s$ in the 3rd term above can increase the $n_s$-value by $1.0\cdot 10^{-3}$, and the subleading contribution to the tensor-to-scalar-ratio $r$, given by
the 3rd term above, can decrease the $r$-value by $5\cdot 10^{-4}$. The subleading contributions are within the observational errors  given in Eq.~(\ref{cmb}). On the other hand, in the first order with respect to the string parameter $\g$, the GZ quantum contributions to $n_s$ and $r$ are up to $+2.5\cdot 10^{-4}$ and $-4.9\cdot 10^{-5}$, respectively. 

Therefore, the GZ quantum contributions to the CMB tilts are smaller than the subleading terms proportional to $N^{-2}_*$ by one order
of magnitude but may be of the same order of magnitude as the classical $N^{-3}_*$ contributions. The same conclusion also applies to
the running parameters $\a_s$ and $\a_t$.

\section{Universal reheating after Starobinsky inflation}

Starobinsky inflation also implies the {\it universal reheating mechanism}, as an additional bonus, 
when matter is coupled to Starobinsky gravity in the Jordan frame \cite{Gorbunov:2010bn}. Reheating was a transfer of energy from inflaton to ordinary particles when inflaton field coherently oscillated around the minimum of its potential in an expanding universe. Reheating took place after inflation but before the Big Bang Nucleosynthesis (BBN) and radiation domination. Before reheating, the leading channel of particle production can be due to a nonperturbative (broad) parametric resonance, known as preheating that is not considered here (see, however, Ref.~\cite{Ketov:2012se} about preheating in the Starobinsky supergravity framework).  The resonance disappeared when the inflaton field became sufficiently small, and it was replaced by perturbative decays. Reheating provided the initial conditions for leptogenesis and baryogenesis,  DM abundance, baryon asymmetry, relic monopoles and gravitinos, and BBN. Both preheating and reheating are highly model-dependent, while Starobinsky gravity coupled to matter gives testable predictions.

The classical solution near the minimum  of the inflaton scalar potential reads
\be \lb{zmin} 
 a(t)\approx a_0\left(\fracmm{t}{t_0}\right)^{2/3}\qquad {\rm and} \qquad
\vf(t)\approx\left( \fracmm{M_{\rm Pl}}{3M} \right)
\fracmm{\cos\[M(t-t_0)]}{t-t_0}~, \ee
where $M$ is the inflaton mass (Sec.~2). The time-dependent classical spacetime background leads to  
quantum production of particles with masses $m<M$. When particle production is included, the amplitude of inflaton 
oscillations decreases much faster via inflaton decay and the universe expansion as
\be \lb{klsa}
\exp[-\ha (3H+\G_{\vf})t]~,
\ee 
where $\G_{\vf}$ is the perturbative decay rate (see below). Accordingly, the inflaton equation of motion gets modified as
\be \lb{ppue}
\ddt{\vf}+(3H+\G_{\vf})\dt{\vf}+V_{,\vf}=0~.
\ee 

The Starobinsky model (in the Jordan frame) with matter $\j$ is described by an action 
\be \lb{starjo}
 S = \int d^4x \sqrt{-g\low{J}}F_{\rm Star.}(R\low{J}) + S_{\rm matter}(g^{\m\n}\low{J},\j)~.
\ee
After the Legendre-Weyl transformation to the Einstein frame it takes the form
\be \lb{starei}
S = S_{\rm quintessence}(g\low{\m\n},\vf) 
+ S_{\rm matter}(g^{\m\n}e^{-\tilde{\k}{\vf}},\j)~,
\ee 
where inflaton $\vf$ couples to {\it all} non-conformal terms and fields $\j$ due to universality of gravitational interaction, and 
$\tilde{\k}\sim M_{\rm Pl}^{-1}$. Therefore, 
Starobinsky inflation leads to the universal mechanism of particle production, see Ref.~\cite{Gorbunov:2010bn} for some specific examples. The decay rates are sensitive to spin and mass of created particles, while all of them are suppressed by the factors of $M_{\rm Pl}$.  By construction, conformal couplings do not contribute to inflaton decay.

The perturbative decay rates of inflaton into a pair of scalars $(s)$ and into a pair of  spin-1/2 fermions $(f)$ are given by \cite{Starobinsky:1981vz,Vilenkin:1985md}
\be \lb{drates}
 \G_{\vf\to ss} =\fracmm{M^3}{192\p M^2_{\rm Pl}} \qquad
{\rm and} \qquad 
\G_{\vf\to ff} =\fracmm{MM^2_f}{48\p M^2_{\rm Pl}}~~, 
\ee
respectively.  The energy transfers by the time $t_{\rm reh} \geq \left(\sum_{s,f}\G_{s,f}\right)^{-1}=\G_{\rm total}^{-1}$. 
One expects quick thermalization after inflation. i.e. a thermal bath filled with baryons, radiation and DM. 
The reheating temperature (the maximal temperature of primordial plasma) after Starobinsky inflation is given by \cite{Starobinsky:1983zz}
\be \lb{rehte}
 T_{\rm reh} \propto \sqrt{ \fracmm{M_{\rm Pl}\G_{\rm total}}{(\# d.o.f.)^{1/2}} }
\approx 10^9 ~{\rm GeV}~, \ee
where the effective \# d.o.f. $(=g)$  for the Standard Model is  $g=106.75$, and $g=228.75$ for the Minimal Supersymmetric Standard Model.  Once the universe is thermalized, the energy density of radiation $\r_{\rm reh}$ is related to the reheating temperature as
 \be \lb{rehd}
 \r_{\rm reh} = \fracmm{\p^2 g}{30} T^4_{\rm reh} ~.
 \ee

The Starobinsky reheating may also lead to production of DM particles and right-handed Majorana neutrinos via scalaron decay, may explain the neutrino oscillations (via the standard seasaw mechanism)  and the baryon asymmetry (via the standard leptogenesis and  the related baryogenesis) also \cite{Gorbunov:2010bn,Jeong:2023zrv}.

The specific predictions of Starobinsky inflation for the cosmological tilts (Secs. 2 and 5) are dependent upon the e-folds number $N_*$ 
that is also affected by the amount of reheating at the end of inflation. A combination of the Planck, BICEP2/Keck Array and BAO data implies $50\leq  N_* \leq 52$ at the 68\% CL \cite{Ellis:2015pla,Ellis:2015jpg}, which is still consistent with (\ref{Nstar}) but also implies  tight  constraints on Starobinsky inflation.

\section{Conclusion}

 The upcoming CMB measurements by CORE Collaboration \cite{CORE:2016ymi}, S4 Collaboration \cite{CMB-S4:2020lpa}, 
 LiteBIRD Collaboration \cite{Paoletti:2022kij}, NASA PICO Collaboration \cite{NASAPICO:2019thw}, the Simons Observatory survey \cite{SimonsObservatory:2018koc} and EUCLID Collaboration  \cite{Euclid:2021qvm} are expected to probe the tensor-to-scalar ratio $r$ in the range of $10^{-3}$ and improve the precision value of $n_s$. Should those measurements confirm Eq.~(\ref{starrel}), this would be a triumph of Starobinsky inflation. Should the disagreement in $r$ be significant (say, two orders of magnitude), it would rule out Starobinsky inflation. Should the disagreement be small (say, within one order of magnitude), one may expect that due to the sub-leading corrections that may be either due to the sub-leading terms in the classical Starobinsky model or due to quantum gravity corrections to the gravitational EFT, or may have a different origin, say, due to reheating or new physics, see e.g., Ref.~\cite{Aldabergenov:2018qhs}.
 
The Starobinsky model is just the simplest large-single-field model of inflation, which has solid theoretical motivation and well fits the current CMB measurements 45 years after the pioneering paper \cite{Starobinsky:1980te}, which is a quite remarkable achievement. There exist no obstruction toward generalizations of single-field models to multi-field models of inflation preferred by supergravity and string theory. Should the primordial  non-Gaussianity be detected, it would be another crucial test of Starobinsky inflation because a contribution of the nonlinear (NL) 3-point correlation function of primordial curvature perturbations to the amplitude of primordial non-Gaussianity is fixed by the Maldacena relation  $f_{\rm NL}=\fracm{5}{12}(1-n_s)$ in all single-field models of inflation \cite{Maldacena:2002vr}.
 
 \section*{Acknowledgements}

The author (SVK) is grateful to Ignatios Antoniadis, Robert Brandenberger, Michael Duff, Gia Dvali and Dmitry Gorbunov for discussions.  SVK was supported by Tokyo Metropolitan University, the Japanese Society for Promotion of Science under the grant No.~22K03624, the World Premier International Research Center Initiative (MEXT, Japan), and Tomsk State University. 

\bibliographystyle{utphys}
\bibliography{references}

\end{document}